\documentclass[12pt]{article}

\usepackage{cite}
\usepackage{graphicx}
\usepackage{epsfig}
\usepackage{amsfonts}
\usepackage{amssymb}
\usepackage{amsmath,amssymb}
\usepackage{color}
\usepackage{epsf,epsfig}
\usepackage{amsmath}
\usepackage{amssymb}
\usepackage{mathrsfs}

\date{\today}

\newcommand{\pont}{{\,^\ast\!}R\,R}

\newcommand{\ee}{\end{equation}}
\newcommand{\eea}{\end{eqnarray}}
\newcommand{\be}{\begin{equation}}
\newcommand{\bea}{\begin{eqnarray}}

\begin{document}
\begin{center}

{\Large \bf 
Non-perturbative spinning black holes
\\
 in dynamical Chern-Simons gravity
} 
\vspace{0.6cm}
\\  
{\large Terence Delsate}$^{1}$,  
{\large Carlos Herdeiro}$^{2}$,  
and
{\large Eugen Radu}$^{2}$
\vspace{0.3cm}
\\  
$^{1}${\small Physique-Math\'ematique, Universite de
Mons-Hainaut, Mons, Belgium}
\\
$^{2}${\small Departamento de F\'\i sica da Universidade de Aveiro  and \\ Centre for Research and Development in Mathematics  and Applications (CIDMA)} 
\\
{\small Campus de Santiago, 3810-183 Aveiro, Portugal} 
\end{center}

\small{
\begin{abstract}
Spinning black holes in dynamical Einstein-Chern-Simons gravity are constructed by directly solving the field equations, without resorting to any perturbative expansion. This model is obtained by adding to the Einstein-Hilbert action a particular higher-curvature correction: the Pontryagin density, linearly coupled to a scalar field. 
The spinning black holes are stationary, axi-symmetric, asymptotically flat generalisations of the Kerr solution of Einstein's gravity, but they possess a non-trivial (odd-parity) scalar field. 
They are regular on and outside the horizon and satisfy a generalized Smarr relation. 
We discuss the deviations from Kerr at the level of the spin and mass distribution, the horizon angular velocity, the ergo-region and some basic properties of geodesic motion. For sufficiently small values of the Chern-Simons coupling our results match those previously obtained using a perturbative approach.
\end{abstract}
}

\date{\today}

\section{Introduction}
Amongst all possible four dimensional extensions of General Relativity (GR) containing higher curvature corrections  \cite{Berti:2015itd}, 
the cases of Gauss-Bonnet (GB) and Chern-Simons (CS) models are of special interest.
Both models rely on coupling topological terms to 
matter fields -- in the simplest case, a real scalar field --,
making the (otherwise) topological terms dynamical and contributing to the field equations.
These extra terms, quadratic in the curvature, lead to new effects in the strong-field regime, manifesting themselves most naturally in the BH solutions of these models. 
 
There are, however, important differences between the (scalar-) GB and CS extensions of  GR.
Firstly, whereas the GB term affects the properties of solutions already in the static sector,
 the CS term leads to different results 
only in the presence of a parity-odd source, such as rotation. Thus, any static solution of GR is also a solution
of dynamical Einstein-Chern-Simons (ECS) gravity \cite{Jackiw:2003pm}.
Secondly, while the fully nonlinear 
generalization of the astrophysically relevant 
Kerr metric in Einstein-Gauss-Bonnet (EGB) gravity has been constructed~\cite{Kleihaus:2011tg,Kleihaus:2015aje}, so far only partial (perturbative) results have been known in the ECS case\footnote{The only non-perturbative results are those in~\cite{Brihaye:2016lsx}
for the (more academic) Taub-NUT solution.},
see $e.g.$~\cite{Yunes:2009hc,Konno:2009kg,Cambiaso:2010un,Cambiaso:2010cg,Yagi:2012ya,Stein:2014xba,Konno:2014qua,McNees:2015srl,Maselli:2017kic}.
The absence of a fully nonlinear version of the Kerr metric in ECS gravity
is  presumably due to the complexity of the problem, since the equations of motion
contain third order order derivatives~\cite{Delsate:2014hba}; by contrast, they remain second order in  EGB gravity, since the GB invariant is an Euler density, and Euler densities are the building blocks of the most general metric gravitational theory with second order field equations, $i.e.$ Lovelock gravity~\cite{Lovelock:1971yv}.  
Still, ECS gravity remains physically interesting, both as an example of parity violating gravity and due to its motivation in quantum gravity approaches, such as string theory~\cite{Campbell:1990fu} and loop quantum gravity~\cite{Taveras:2008yf,Ashtekar:1988sw} - see~\cite{Alexander:2009tp} for a review.

The perturbative construction of BH solutions in ECS gravity,  
as in~\cite{Yunes:2009hc}-\cite{McNees:2015srl}, 
has the advantage of yielding closed form expressions, leading to insights on the trends introduced by the CS coupling.
It is clear, however, that a number of important features, occurring in the fast spinning and/or large coupling regimes,  cannot be captured by  this perturbative approach.  The main purpose of this work is to present a general framework which allows a
non-perturbative approach to constructing the modified Kerr solution in ECS model, together with some numerical results
 illustrating non-perturbative solutions, including fast rotating BHs.
 
 This paper is organised as follows. In Section~\ref{sec2} we present the model, including the equations of motion, boundary conditions and relevant physical quantities. Some details on the numerical approach are also presented. The results are presented in Section~\ref{sec3}, where we comment both on the properties of illustrative solutions and on an overview of the solution's properties as one deviates from the GR limit. Final remarks are presented in Section~\ref{discussion}. We use units with $c=1$.

\section{The model}
\label{sec2}
  
\subsection{Action, equations of motion and ansatz}

A general  ECS gravity model is described  by the action
\begin{eqnarray}
\label{CSaction}
\mathcal{S}=\int d^4x  \sqrt{-g} 
\left[
 \frac{R}{16 \pi G} 
+ \frac{\alpha}{4} f(\phi)
{{\,^\ast\!}R\,R} 
 -\frac{1}{2} g^{ab} (\nabla_a \phi) (\nabla_b \phi) 
-V(\phi)
\right] ~,
\end{eqnarray}
where $\phi$ is a real scalar field 
with a potential $V(\phi)$, 
$f(\phi)$ is a coupling function  and $\alpha$ a dimensionful constant, an input parameter of the theory.
As usual,
$g$ is the determinant of the metric $g_{\mu\nu}$
and
$R$ is the Ricci scalar.
 The quantity $\pont$ is the Pontryagin density, also referred to as the Chern-Simons scalar, defined via\footnote{Note that the Pontryagin term is topological and
 can be expressed as a total divergence \cite{Alexander:2009tp}
\be
\nabla_a K^a = \frac{1}{2} \pont \ , \qquad {\rm with}~~
K^a \equiv \epsilon^{abcd} \Gamma^n_{bm} \left(\partial_c\Gamma^m_{dn}+\frac{2}{3} \Gamma^m_{cl}\Gamma^l_{dn}\right)\, .
\label{eq:curr1}
\ee
}
\begin{eqnarray}
\label{pontryagindef}
\pont=  
 {\,^\ast\!}R^a{}_b{}^{cd} R^b{}_{acd}~, \qquad ~~{\rm with}
~~{^\ast}R^a{}_b{}^{cd}\equiv \frac12 \epsilon^{cdef}R^a{}_{bef}~,
\end{eqnarray}
where $\epsilon^{cdef}$ is the 4-dimensional Levi-Civita tensor. 
 
The gravitational equations derived from~\eqref{CSaction} can be written in a GR-like form:
\begin{eqnarray}
\label{ECSeqs}
E_{ab}\equiv G_{ab}-8 \pi G ~T_{ab} =0\ ,~ 
\end{eqnarray}
where $G_{ab}$ is  the Einstein tensor and the \textit{effective} energy momentum tensor 
\begin{eqnarray}
\label{Teff}
T_{ab} \equiv  T_{ab}^{(\phi)}-2 \alpha  C_{ab}\  ,
\end{eqnarray}
is a combination of the scalar field energy-momentum tensor, $T_{ab}^{(\phi)} $,  
\be
\label{Tab-theta}
T_{ab}^{(\phi)}
\equiv    \left(\nabla_{a} \phi \right) \left(\nabla_{b} \phi \right) 
    - g_{a b}\left [ \frac{1}{2}  \left(\nabla_{c} \phi \right) \left(\nabla^{c} \phi\right) +  V(\phi) \right] \ ,
\ee
and a contribution from the CS term,
\begin{eqnarray}
\label{Ctensor}
C^{ab} = (\nabla_c f(\phi))
\epsilon^{cde(a}\nabla_e R^{b)}{}_d+
(\nabla_{c}\nabla_{d} f(\phi))
{\,^\ast\!}R^{d(ab)c}\ .
\end{eqnarray}
The scalar field equation derived from~\eqref{CSaction} is a modified Klein-Gordon equation,
\be 
\label{KGeq}
\nabla^2 \phi = \frac{dV}{d\phi} - \frac{\alpha}{4} \frac{df(\phi)}{d\phi} \pont \ .
\ee

To make contact with previous literature, in this work we shall report results for a massless, non-self-interacting scalar field, 
and a linear coupling function
\begin{eqnarray}
\label{choice}
 V(\phi)=0 \ ,~~~{\rm and}~~~~  f(\phi)=\phi \ .
\end{eqnarray}
Some brief comments on more general cases will be made in Section~\ref{discussion}.

To obtain stationary and
axi-symmetric BH spacetimes, possessing
two commuting Killing vector fields, $\xi$ and $\eta$, we use a coordinate system adapted to these symmetries. Then: 
$
\xi = \partial_t,
$
$
\eta=\partial_\varphi.
$
Such spacetimes are usually described by a Lewis-Papapetrou-type
ansatz~\cite{Wald:rg}, which contains four unknown functions.
We shall use here a metric ansatz originally introduced in \cite{Herdeiro:2014goa,Herdeiro:2015gia},
which factorizes the asymptotics
\begin{eqnarray}
\label{metric}
ds^2 =-e^{2F_0(r,\theta)} N dt^2+e^{2F_1(r,\theta)}\left(\frac{dr^2}{N}+r^2 d\theta^2\right)+e^{2F_2(r,\theta)}r^2 \sin^2\theta (d\varphi-W(r,\theta) dt)^2
\ , 
\end{eqnarray} 
where $N\equiv1-{r_H}/{r}$, and $r_H$ is a constant. The scalar field $\phi$ depends on the $r,\theta$ coordinates only:
%
%
 \begin{eqnarray}
 \label{scalar}
 \phi \equiv \phi(r,\theta)~.
 \end{eqnarray}

 \subsection{Boundary conditions and physical quantities}

We are interested in asymptotically flat  solutions.
 This implies the following boundary conditions\footnote{
Setting $\phi\big |_{r=\infty}=0$ removes the shift  symmetry 
 $\phi\to \phi+const$.
of the ECS model~\eqref{CSaction} with (\ref{choice}).}
\begin{eqnarray}
\label{infinity}
 F_i\big |_{r=\infty}=W\big |_{r=\infty}=\phi\big |_{r=\infty}=0 \ ,
 \end{eqnarray}
where $i=0,1,2$. Since the scalar field is massless, one can construct an approximate solution of the field equations  
compatible with these asymptotics as a power series in $1/r$. The leading order terms of such an expansion
are:
\begin{eqnarray}
\label{r-infty}
\nonumber
&&
F_0(r,\theta)=\frac{c_t}{r} 
+\dots, \qquad
F_1(r,\theta)=-\frac{c_t}{r} 
+\dots, \qquad
F_2(r,\theta)=-\frac{c_t}{r} 
+\dots, \nonumber \\
&&
W(r,\theta)=\frac{c_\varphi}{r^3} 
+\dots, \qquad
\phi(r,\theta)=\frac{q \cos \theta }{r^2 }+..,
\end{eqnarray} 
where $c_t$, $c_\varphi$ and $q$ are constants fixed by the numerics;  $q$ corresponds to the dipole moment of the scalar field, which has no monopole term.

On the
symmetry axis, $i.e.$ at $\theta=0,\pi$,  axi-symmetry and regularity require that
\begin{eqnarray}
\label{axis}
\partial_\theta F_i\big |_{\theta=0,\pi}
=  \partial_\theta  W\big |_{\theta=0,\pi}=\partial_\theta \phi\big |_{\theta=0,\pi}=0\ ,~
 \end{eqnarray}
Again,  an approximate expansion of the solution compatible with these conditions can be constructed;
at, say, $\theta=0$ one finds
\begin{eqnarray}
\label{t0}
{\cal F}_a(r,\theta)= {\cal F}_{a0}(r)+\theta^2 {\cal F}_{a2}(r)+\mathcal{O}(\theta^4)\ ,
\end{eqnarray} 
where ${\cal F}_a =\{F_0, F_1, F_2, W; \phi\}$, and
the essential data, which is fixed by the numerics, is encoded in the 
functions ${\cal F}_{a0}=\{F_{i0},W_{0},\phi_{0}\}$.  Absence of conical singularities requires, moreover, 
$F_1\big |_{\theta=0,\pi}=F_2\big |_{\theta=0,\pi}$. 

For the considered coupling function~\eqref{choice}, 
the problem possesses a well defined parity:
the metric functions are invariant under a reflection along the equatorial plane $\theta=\pi/2$,
while the scalar field changes sign, $\phi \to -\phi$.
This symmetry is used to integrate the field equations for $0\leqslant \theta \leqslant \pi/2$ only.

For the metric ansatz~\eqref{metric}, the event horizon is located at a surface with constant radial variable, $r=r_H>0$.
By introducing a new radial coordinate $x\equiv \sqrt{r^2-r_H^2}$, 
the horizon boundary conditions and numerical treatment of the problem are simplified. These boundary conditions are 
\begin{equation}
\partial_x F_i \big|_{x=0}= \partial_x \phi  \big|_{x=0} =  0\ , \qquad W \big|_{x=0}=\Omega_H\ ,
\label{bch1}
\end{equation}
where $\Omega_H $ is the horizon angular velocity, such  that
the Killing vector $\chi =\xi+\Omega_H \eta$ is orthogonal and null on the horizon.
These conditions are consistent with the near horizon solution:\footnote{A similar near horizon
expansion for the
 Kerr solution in a generic EGB model implies the existence of a critical set of limiting configurations.
That is, the second order term $\phi_2$ solves a quadratic equation; the vanishing of this equation's discriminant selects solutions that form a part of the boundary of the domain of existence.
No such restriction is found for the ECS model.
}
\begin{eqnarray}
\label{rh}
{\cal F}_a(r,\theta)= {\cal F}_{a0}\theta)+x^2 {\cal F}_{a2}(\theta)+\mathcal{O}(x^4)\ ,
\end{eqnarray}  
where the essential functions are
${\cal F}_{i0}$  
(also $F_0\big |_{r_H}=F_1\big |_{r_H}$).

The ADM  mass $M$ and the total angular momentum $J$ of the solutions
are read off from the asymptotics of the metric functions,
\begin{eqnarray}
\label{asym}
g_{tt} =-1+\frac{2GM}{r}+\dots \ , \qquad ~~g_{\varphi t}=-\frac{2GJ}{r}\sin^2\theta+\dots \ .
\end{eqnarray}
As usual (see, $e.g.$,~\cite{Townsend:1997ku}),
$M$ and $J$ can be split 
into the horizon contribution, respectively $M_H$ and $J_H$, -- computed as a Komar integrals on the horizon -- and the ``matter" contribution, respectively, $M_\psi$ and $J_\psi$, in this case composed by the scalar field and CS parts. The latter are  
computed as volume integrals of the appropriate energy-momentum tensor components:
\begin{eqnarray}
\label{TotalMass}
&&
M = M_H+M_\psi\ , \qquad {\rm with}~~
 M_\psi=-2\int_\Sigma dS_a\bigg( T_{b}^{\ a}  \xi^{b}-\frac{1}{2}T \xi^a \bigg)\ ,
\\
\label{TotalAngularMomentum}
 &&
J = J_H +J_\psi\ , \qquad {\rm with}~~
 J_\psi= \int_\Sigma dS_a \left( T_{b}^{a}  \eta^{b} -\frac{1}{2}T  \eta^{a} \right) \ ,
\end{eqnarray}
where $\Sigma$ is a spacelike surface, bounded by the sphere at infinity
$S^2_\infty$ and the horizon $\mathscr{H}$.
In the above relations, $M_\psi$ and $J_\psi$
encode the contribution of the \textit{effective} ``matter"  distribution to the total
mass and angular momentum.
For Kerr BHs, $M=M_H$ and $J=J_H$;  
this is not so for ECS BHs.  
Moreover,
since $T_t^{t(\phi)}-\frac{1}{2}T^{(\phi)}=T_\varphi^{t(\phi)}=0$,
only the CS part of the \textit{effective} energy-momentum tensor (\ref{Teff})
contributes to the 
energy and angular momentum ``matter" densities,
which are determined by the
 $C_{t}^t$ and  $C_{\varphi}^t$ components, respectively (since $C_a^a=0$).

The BH horizon introduces a temperature $T_H$ and horizon area $ A_H $,
\begin{eqnarray}
\label{THAH}
T_H=\frac{1}{4\pi r_H}e^{(F_0-F_1)|_{r_H}}\ ,
\qquad
A_H=2\pi r_H^2 \int_0^\pi d\theta \sin \theta~e^{(F_1+F_2)|_{r_H}} \ .
\end{eqnarray}
Then a straightforward computation shows that
the following Smarr relation holds in ECS theory:
\begin{eqnarray}
\label{Smarr}
M+{\mathcal U}=2(\Omega_H J+T_H S)\ ,
\end{eqnarray}
where
\begin{eqnarray}
\label{U}
{\mathcal U}=2\alpha  \int_\Sigma  dS_a \xi^a (\nabla \phi)^2\ ,
\end{eqnarray}
and $S$ is the BH entropy in the ECS model, which is the sum of two different contributions 
\cite{McNees:2015srl}
\begin{eqnarray}
\label{S}
S=S_E+S_{CS}\ , \qquad
~~{\rm with}~~S_E\equiv \frac{A_H}{4G}\ , \qquad 
S_{CS}\equiv \pi \alpha \int_\mathscr{H}  \phi 
{\,^\ast\!}R^{abcd} \hat \epsilon_{ab}  \hat \epsilon_{cd}  \hat \epsilon~,
\end{eqnarray}
  $ \hat \epsilon_{ab}$ being the binormal of the horizon,
which is normalized such that $ \hat \epsilon_{ab}  \hat\epsilon^{ab}=-2$.
One can also show that the following relation holds
\begin{eqnarray} 
T_H S_{CS}=-\alpha \int \hat \epsilon \phi K^r\ . 
\end{eqnarray}

\subsection{Scaling and dimensionless quantities}
The dependence on Newton's constant $G$ 
 disappears from the field equations under the rescaling
\begin{eqnarray}
\label{s1}
\phi\to  \phi/\sqrt{8\pi G}\ , \qquad \alpha \to  \alpha /\sqrt{8\pi G}\ .
\end{eqnarray}
This rescaling makes the scalar field dimensionless, whereas $\alpha$
remains a dimensionful constant.  The field equations still possess the scaling symmetry
\begin{eqnarray}
\label{scaling}
 r\to \lambda r\ , \qquad  \alpha \to  \lambda^2  \alpha \ ,
\end{eqnarray}
where $\lambda$ is a positive constant, under which global quantities transform as  $M\to \lambda M$, $J\to \lambda^2 J$.
In the following we shall work with dimensionless quantities
which are invariant under (\ref{scaling}):
\begin{eqnarray}
\label{xi}
 \xi\equiv \frac{ \alpha \sqrt{8\pi G}}{M^2} \ , \qquad 
j\equiv \frac{J}{M^2}\ , \qquad w_H\equiv \Omega_H M \ .
\end{eqnarray}

\subsection{The numerical scheme}

Within our approach,
the ECS equations reduce to a  system of five 
coupled non-linear elliptic partial differential equations\footnote{These equations are long,
each of them containing several hundred independent terms.} 
for the functions 
${\cal F}_i$. 
These equations consist of 
the Klein-Gordon equation (\ref{KGeq})
together with suitable combinations of the ECS equations (\ref{ECSeqs})
$
\{ 
E_r^r+E_\theta^\theta=0;
E_\varphi^\varphi=0;
E_t^t=0;
E_\varphi^t=0
 \}.
$

The resulting equations, however, contain
third
order derivatives of the metric functions $F_0,F_2$ and $W$. 
To obtain a standard form of the problem,
we introduce  a set of `auxiliary' functions $S_i,Q_i$,
with
\begin{eqnarray}
S_1= F_{0,r}\ ,~S_2=F_{2,r}\ ,~S_3=W_{,r} \qquad 
~{\rm and}~~ \qquad 
Q_1= F_{0,\theta}\ ,~Q_2=F_{2,\theta}\ ,~Q_3=W_{,\theta}\ .
\end{eqnarray}
These `auxiliary' functions satisfy the following boundary conditions
\begin{equation}
S_i\big |_{r=\infty}= Q_i\big |_{r=\infty}=0\ , \qquad 
 \partial_\theta S_i\big |_{\theta=0,\pi/2}= \partial_\theta Q_i\big |_{\theta=0,\pi/2}=0\ , \qquad 
 \partial_x S_i\big |_{x=0}=\partial_x  Q_i\big |_{x=0}=0\ ,
\end{equation}
which are compatible with the approximate expression of the solutions given above.

The 
remaining equations $E_\theta^r =0,~E_r^r-E_\theta^\theta  =0$,
yields two constraints  which are monitored in numerics.
Typically they are satisfied at the level of the overall numerical accuracy.

Our numerical treatment can be summarized as follows: 
$(i)$ we use the radial variable  
$x$ introduced above; 
$(ii)$ this coordinate
 is compactified,
$
 \bar x\equiv {x}/({1+x}).
$
This transformation maps the semi infinite $x$-domain $[0,\infty)$ to the finite $\bar x$-domain $[0,1]$;  
$(iii)$ the equations for ${\cal F}$
are discretized on some given grid in $\bar x$ and $\theta$. 
Various grid choices have been considered, but 
most of the results have been obtained for  
 an equidistant grid with $150 \times 30$ points. 
The grid covers the integration region
$0\leqslant \bar x \leqslant 1$ and $0\leqslant \theta \leqslant \pi/2$; 
$(iv)$
all numerical calculations
are performed with a professional package \cite{schoen}, 
which uses a Newton-Raphson method.
We remark that  for all solutions obtained we
have monitored the Ricci and Kretschmann scalars, and, at the level of the numerical accuracy, we have
not observed any sign of a singular behaviour.  
As a further test, we have verified that
 our results for small $\xi$ and $j$ are in good agreement 
with the perturbative results in~\cite{Yagi:2012ya}.

In this scheme, there are three input parameters: ${\bf i)}$ the event horizon velocity $\Omega_H$;
 ${\bf ii)}$ the event horizon radius $r_H$ in the metric ansatz (\ref{metric}); 
 ${\bf iii)}$ the coupling constant $\alpha$.
The first two parameters are geometric quantities, while the third one characterizes the theory.

In our approach we start with
an Einstein gravity Kerr solution with given $r_H,\Omega_H$ as initial guess\footnote{In practice,
the scaling symmetry (\ref{scaling})
is used to fix the value 
of the event horizon radius.}
for an ECS BH with a small value of $\alpha$.
Then we increase
the value of $\alpha$ slowly.
 The iterations converge, and, in principle, repeating the procedure we obtain in this
way solutions for increasingly higher values of $\alpha$. 
Around one thousand different solutions 
 were constructed in this way,
covering a part of the domain of existence of ECS BHs. 
 
In contrast to the EGB case \cite{Kleihaus:2015aje}, 
no clear upper bound seems to exist on the value of $\alpha$, or equivalently on the dimensionless parameter $\xi$.
For any initial Kerr configuration, however, 
the numerical accuracy decreases with increasing
$\alpha$, the convergence of the numerical iteration becoming slower and
requiring a very large number of iterations (or even being lost).
As such,
we could not scan the full domain of existence
of the solutions.
Very likely, this is a numerical problem only;
we suspect that
a better approach could show that,
for a given $j$ (or $w_H$)
 no upper bound exists for
the parameter $\xi$.

\begin{figure}[ht!]
\begin{center}
\includegraphics[height=.26\textheight, angle =0]{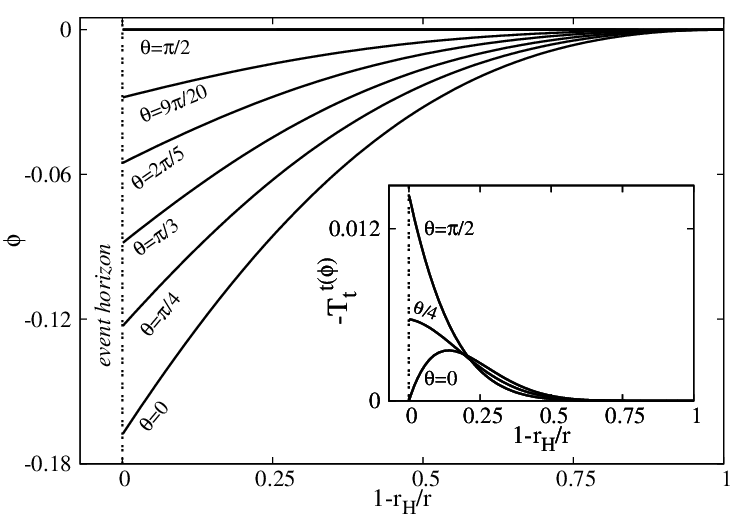} 
\includegraphics[height=.26\textheight, angle =0]{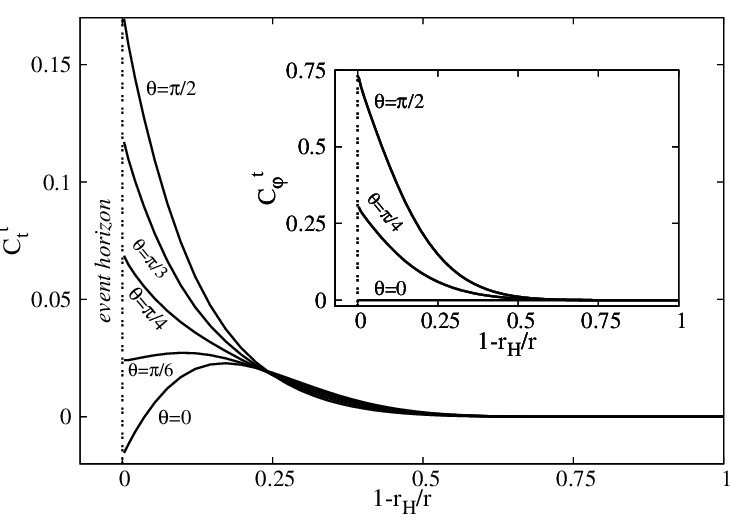}  
\caption{
(Left panel) 
Radial variation of the scalar field $\phi$ and of the $T_t^{t(\phi)}$-component of its energy-momentum tensor for several values of $\theta$. 
(Right panel) 
Same for the $C_t^{t}$ and $C_\varphi^{t}$-components of the \textit{effective} energy-momentum tensor associated
with the energy and angular momentum densities.
The  
 ECS BH in this plot has the input parameters $r_H=1$, $\Omega_H=0.2$ and $\alpha=0.3$,
	while $\xi=0.951$ and $j=0.493$.
}
\label{ZC}
\end{center}
\end{figure} 

\section{Numerical results}
\label{sec3}

\subsection{General properties}

From the results of the numerical integration we have observed that a Kerr BH solution for any $j$ allows a generalization in the ECS model. As expected, the deviation from the GR solution increases with the value of 
the coupling constant 
$\xi\sim \alpha/M^2$.
Since we were not been able to identify the existence of 
an upper bound for $\xi$, neither at the numerical nor at the analytical level, 
for all curves displayed in this work,
the end points correspond to configurations where the numerical results 
stopped being reliable, rather than some fundamental obstruction.\footnote{The typical numerical error
for the solutions displayed in this work is estimated of the order of $10^{-3}$ or lower,
except close to the end points of the curves, where it increases to around a few percent.}

Let us first comment on features of an illustrative solution. Unlike Kerr, ECS spinning BHs have a non-trivial scalar field profile outside the horizon. A typical $\phi$ profile is  shown in Figure \ref{ZC} (main left panel). The northern hemisphere scalar field is negative and a monotonically increasing function of the radial coordinate for each $\theta\neq 0$. At the equator it vanishes (as required) and in the southern hemisphere it changes sign (not shown). This scalar field profile is distinct from those observed in other cases of asymptotically flat spinning BHs with scalar hair, namely those in EGB gravity and Kerr BHs with synchronised scalar hair in GR; in both these cases a typical scalar field profile is not monotonic - see Figure~2 in~\cite{Kleihaus:2015aje} and Figure~16 in~\cite{Kleihaus:2015aje}. A discussion of the energy/angular momentum distribution of the scalar field for the ECS BHs is delicate, since the most meaningful energy/angular momentum densities, $i.e.$ those entering eqs.~\eqref{TotalMass} and \eqref{TotalAngularMomentum} vanish. Still, some intuition may, perhaps, be gained from the inset of Figure \ref{ZC} (left panel) where we exhibit only the time-time component of the scalar field energy momentum tensor. Such energy-density is asymptotically vanishing; since it also vanishes at the BH poles, it attains a maximum along the symmetry axis at some distance from the horizon. The right panel of Figure \ref{ZC} shows the analogous quantity for the CS term contribution to the effective energy momentum tensor (main panel), where a similar behaviour is observed. The corresponding angular momentum density is shown in the inset.

\begin{figure}[ht!]
\begin{center}
{\includegraphics[width=9cm]{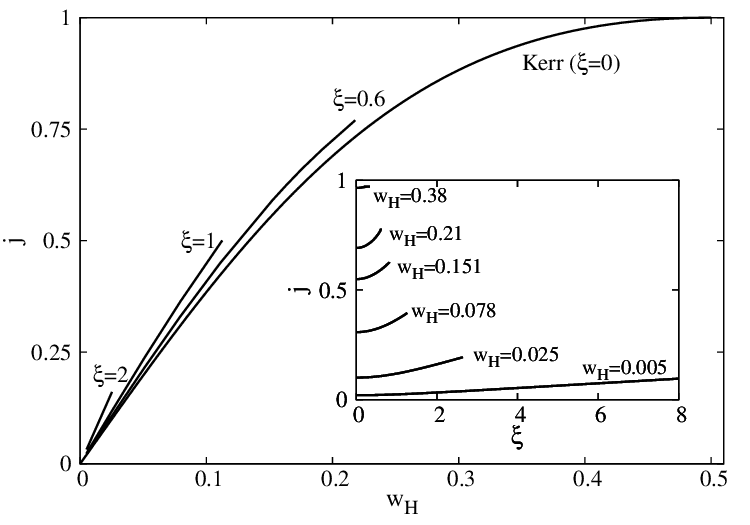}}  
\caption{The $j(w_H)$-diagram of ECS BHs, for several values of the dimensionless coupling constant $\xi$.
(Inset) The $j(\xi)$ curves  
 for different values of $w_H$.
\label{jwH}
}
\end{center}
\end{figure}

We now turn our attention to trends in the space of solutions. In Figure \ref{jwH}
we show the dimensionless spin $j$ as a function of the dimensionless horizon angular velocity $w_H$ for several values of $\xi$. 
One observes that the GR pattern is shared by the ECS solutions: $j$ and $w_H$ are positively correlated, but increasing $\xi$ the same $w_H$ requires an increasingly larger $j$.  Such increase of $j$  with $\xi$, for fixed $w_H$ is detailed in the inset (for Kerr BHs
$j=4w_H/(1+4w_H^2)$).   This confirms that ECS spinning BHs require more (dimensionless) angular momentum to support the same (dimensionless) angular velocity as Kerr BHs. This property is likely related to the fact that not all angular momentum is stored inside the  horizon for ECS BHs.
We emphasize that we have constructed 
ECS BHs with $0<j<1$,
although the near-extremal configurations\footnote{
Only non-extremal solutions are reported here.
Extremal ECS BHs should, however, also exist~\cite{Chen:2018jed}.
In contrast to the EGBs case~\cite{Kleihaus:2015aje},
the contribution of the CS term is compatible with a non-singular extremal horizon.}  were found for small values of $\xi$ only.
All ECS spinning BHs constructed so far obey the Kerr bounds for dimensionless spin, $j<1$, 
and dimensionless angular velocity, $w_H<1/2$.

\begin{figure}[ht!]
\begin{center}
\includegraphics[height=.26\textheight, angle =0]{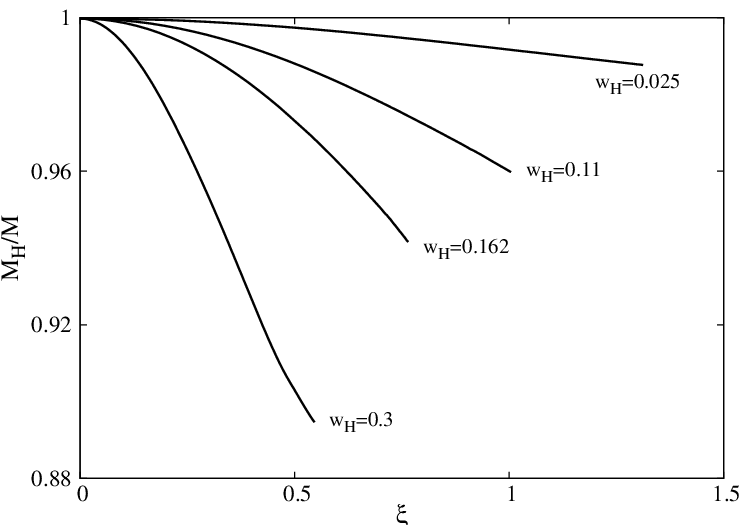} 
\includegraphics[height=.26\textheight, angle =0]{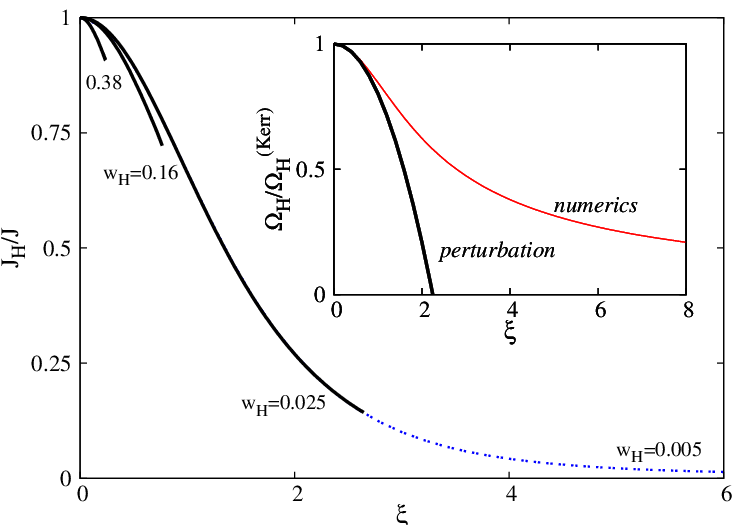}  
\caption{Ratio $M_H/M$ (left panel) and $J_H/J$ (right panel) as a function of  $\xi$
for several values of $w_H$.
(Inset) Ratio $\Omega_H/\Omega_H^{(Kerr)}$ as a function of $\xi$
for BHs with the same $M,J$. 
\label{ratioJ}
}
\end{center}
\end{figure}

To gain some insight on how the mass and angular momentum of ECS BHs are distributed in and outside the horizon, in Figure \ref{ratioJ} we display the ratios
$M_H/M$ and
$J_H/J$.
One can see that most of the mass is stored inside the horizon,
 the ratio
$M_H/M$
decreasing with both $w_H$ and $\xi$. 
By contrast,  
for small $w_H$ and sufficiently large $\xi$, 
most of the angular momentum 
can be stored
\textit{outside} the
horizon. From the data sample analysed, however, we found no evidence of a counterrotating horizon, $J_H/J<0$, as suggested by perturbative studies.

 A somewhat unexpected result concerns 
the ratio $\Omega_H/\Omega_H^{(Kerr)}$ 
for ECS and GR  BHs with the same mass and angular momentum - Figure~\ref{ratioJ} (right panel inset).
\textit{A priori},  the value of this ratio will depend both on $\xi$ and $j$;
however, our results show that the $j$-dependence is weak for all solutions constructed so far; consequently, the same curve (red thin line in the inset of Figure \ref{ratioJ})  provides a  good fit for all data, regardless of $j$. The same panel also shows the perturbation theory result~\cite{Yagi:2012ya}
$
\Omega_H/\Omega_H^{(Kerr)}=1-\frac{709}{3584} \xi^2~
$; one can observe it provides a good approximation up to $\xi$ of order one.

We have also
verified that the ECS BHs are (generically) algebraically general (Petrov type I).
Also, since the metric functions  are always smooth and finite outside the horizon,
the Lorentzian signature of the metric is preserved there. Moreover, 
in all dataset analysed we observed the absence of closed causal  curves.

\subsection{Other features}

All ECS BHs  have an ergoregion, defined as the domain in which $\xi=\partial_t$ is positive (exterior to the horizon). 
This region is bounded by the event horizon and by the surface where
\begin{equation}
 g_{tt}=-e^{2F_0} N+W^2e^{2F_2}r^2 \sin^2\theta =0 \ .
\end{equation}
For the Kerr spacetime, this surface has a spherical topology and touches the horizon at the poles. 
As
discussed in\cite{Herdeiro:2014jaa}, the ergoregion can be more complicated for BHs with scalar hair, with the possible
existence of an additional $S^1\times S^1$ ergo-surface (ergo-torus). We have found that this is not the
case for ECS BHs, where all solutions are Kerr-like and possess a single topologically $S^2$ ergosurface.

\begin{figure}[ht!]
\begin{center}
{\includegraphics[width=9cm]{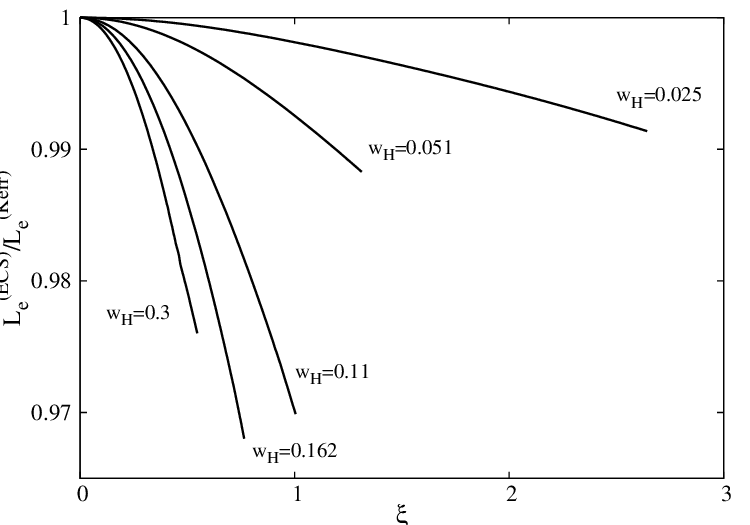}}  
\caption{Ratio between the ECS ergocircle
and the corresponding value for a Kerr BH with the same $(M,J)$
as a function
 of  $\xi$
 for several values of $w_H$. 
\label{ergo}
}
\end{center}
\end{figure}

The effect of the CS term on the ``size" of the ergoregion is illustrated in Figure \ref{ergo}, by using the measure $L_e$, the proper length of the ergocircle along the equatorial plane:
\begin{eqnarray}
L_e^{(ECS)}=r_e^2e^{F_2(r_e,\pi/2)}\ , \qquad ~~{\rm with}~~g_{tt}(r_e,\pi/2)=0\ .
\end{eqnarray}
We have denoted as $L_e^{(Kerr)}$  the corresponding value for a Kerr BH with the same mass and angular momentum.
One can see that the generic effect of the CS term is to reduce the size of the ergoregion as compared to the GR case.  Although increasing $\xi$ one observes a higher value of $j$ for the same $\omega_H$, $cf.$ Figure~\ref{jwH}, the angular momentum is stored both in and outside the horizon, thus making the ergoregion grow less than for a comparable Kerr BH. A similar situation has been observed for Kerr BHs with synchronised hair, which has the physical impact of making superradiant instabilities weaker for the hairy BHs than for comparable Kerr BHs~\cite{Herdeiro:2014jaa,Ganchev:2017uuo,Degollado:2018ypf}.

Let us also briefly look at some basic features of geodesic motion. It is known that the Kerr spacetime supports unstable photon orbits
with a fixed radial coordinate~\cite{Teo,Cunha:2017eoe}.
A subset of the
latter are restricted to the equatorial plane $\theta=\pi/2$, and comprises two independent circular
photon orbits with opposite rotation directions, usually dubbed as {\it light rings}.

\begin{figure}[ht!]
\begin{center}
\includegraphics[height=.26\textheight, angle =0]{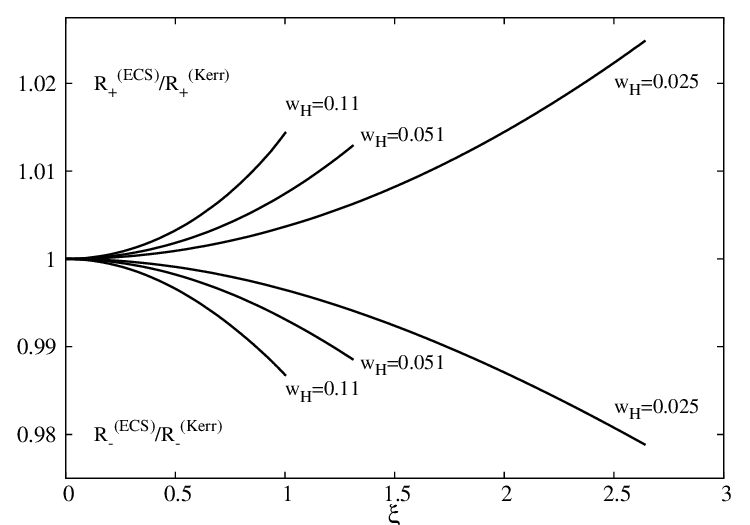} 
\includegraphics[height=.26\textheight, angle =0]{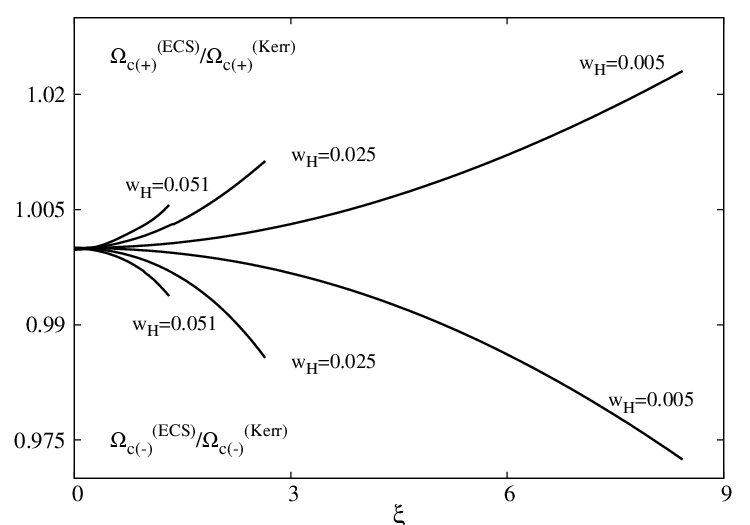}  
\caption{
(Left panel) 
Ratio between the ECS light ring proper radius
and the corresponding value for a Kerr BH with the same $(M,J)$,
as a function
 of the coupling constant $\xi$
 for several  values of $w_H$. 
(Right panel) 
The same for the ratio between the angular frequency at the ISCO.
}
\label{LR1}
\end{center}
\end{figure}

For a stationary spacetime,
the light ring positions can be obtained by analysing the following
condition in the equatorial plane \cite{Cunha:2016bjh}
\begin{eqnarray}
\label{LR}
\partial_rh_\pm=0,\quad \textrm{with}\quad h_\pm=\frac{-g_{t\varphi}\pm\sqrt{g^2_{t\varphi}-g_{tt}g_{\varphi\varphi}}}{g_{tt}}\ . 
\end{eqnarray}
Each sign $\pm $ leads to one of the two light rings.
As shown in Figure \ref{LR1} (left panel),
the light ring qualitative structure is still the same as in Kerr and in the dataset analysed the differences with respect to comparable Kerr BHs are small, at percent level. We thus predict that the BH shadows are going to be very similar to those of Kerr BHs in this region of the solution space. 
We have denoted as $R_{\pm}^{(ECS)}$ the proper length of the light ring orbit,
\begin{eqnarray}
R_{\pm}^{(ECS)}\equiv r_{\pm}e^{F_2(r_{\pm},\pi/2)}, 
\end{eqnarray}
 where $r_\pm$ are the solutions of~Eq. (\ref{LR}).

We have also studied the angular frequency at the ISCO for a large set of ECS BHs.
The geodesic  motion is studied along the equatorial plane, $\theta=\pi/2$; then
the Lagrangian ruling the motion of a timelike test particle (the only case studied here) is
\begin{eqnarray}
2{\cal L}=e^{2F_1 } \frac{\dot r^2}{N }+e^{2F_2 } r^2(\dot \varphi-W \dot t)^2
-e^{2F_0} N \dot t^2=-1\ .
\end{eqnarray}
Note that $F_i,W$ depend only on $r$ for equatorial motions;
also a dot denotes a derivative $w.r.t.$ proper time.
Stationarity and axisymmetry imply the existence of the
 first integrals
\begin{eqnarray}
e^{2F_2}r^2(\dot \varphi-W \dot t)\equiv L\ , \qquad 
~~
(e^{2F_0} N-e^{2F_2} r^2 W^2)\dot t+e^{2F_2} r^2 W \dot \varphi \equiv E\ ,
\end{eqnarray}
where $E$ and  $L$ are
the specific energy and angular momentum of the test particle.
Then the orbital angular velocity is expressed as
\begin{eqnarray}
\Omega_c=\frac{\dot \varphi}{\dot t}=W-\frac{e^{2F_0-2F_2}LN}{r^2(LW-E)}\ .
\end{eqnarray}

The equation governing the variation of the radial coordinate $r$
for an orbit on the equatorial plane is
\begin{eqnarray}
\label{V}
\dot r^2=V(r)=e^{-2F_1}N
\left(
-1-e^{-2F_2}\frac{L^2}{r^2}
+\frac{e^{-2F_0}(E-LW)^2}{N}
\right) \ .
\end{eqnarray}

The requirement for a circular orbit at $r=r_c$ is 
$V(r_c)=V'(r_c)=0$
which
results in two algebraic equations
for $E,L$ which are solved analytically, possessing
two distinct pairs of solutions $(E_{+},L_{+})$ and $(E_{-},L_{-})$,
corresponding to co-rotating and counter-rotating trajectories.

The solutions for $E,L$
are then replaced in the expression of $V''(r_c)$, requiring 
$V''(r_c)\leqslant 0$ for stability.
For the configurations studied so far we have noticed a 
(qualitative) analogy with the Kerr BH.
First, circular geodesic motion is only possible for radii larger than a minimum value, 
$r_c>r_{min}$, a  constraint imposed by 
 requiring the energy $E$
to be real.
Then for $r_{min}<r_c<r_{ISCO}$
only unstable circular orbits can exist, $i.e.$ with
$V''(r_c)> 0$.
For $ r_c>r_{ISCO}$, circular orbits are stable.

In Figure~\ref{LR1} (right panel) we exhibit the angular frequency at the ISCO for co-rotating and counter-rotating geodesics, where 
the value of $\Omega_c$ is normalized with respect to that of a Kerr BH
with equal mass and angular momentum.
Only small deviations from GR were found so far, no larger than a few percent, similarly to the case of the light rings described above. More significant differences are likely to occur for larger values of $\xi$.


\section{Further remarks}
\label{discussion}

The main purpose of this work was to provide a concrete approach to 
constructing the nonperturbative spinning BHs in dynamical ECS gravity,
together with a preliminary discussion of their (basic) physical properties.
These configurations can be viewed
as the counterparts of Kerr solutions in the presence of a CS term in the gravitational action.
Our results here show that such BHs qualitatively share some basic properties of the GR BHs.

The research initiated here can be furthered in many possible directions. 
An important issue is to thoroughly scan the parameter space of solutions and clarify its boundaries and possible limiting configurations.
The stability of the ECS BHs is another important point,
although,
due to the complexity of the field equations,
any result in this direction will be highly challenging task. 
In this context, let us remark that,
as discussed above, these spinning BHs have an ergoregion. 
Thus, similar to the Kerr case, they should be
afflicted by superradiant instabilities in the presence of (massive) bosonic fields~\cite{Brito:2015oca}.
Yet another interesting direction would be to further explore the astrophysical signatures of  these BHs.
An obvious task here will be to study the geodesics in a more systematic way and to compute, $e.g.$, the shadows, or the $X$-ray spectrum in the presence of an accretion disk, contrasting the results with those for the Kerr solution. Phenomenological studies of these features for EGB BHs have revealed only small quantitative differences occur, with respect to Kerr BHs~\cite{Cunha:2016wzk,Zhang:2017unx}. 

It would also be interesting to extend the solutions 
in this work to 
models beyond the simple choices in (\ref{choice}) for $V(\phi)$ and $f(\phi)$.
Working with the same linear scalar coupling to the CS term,
we have constructed families of solutions for a massive scalar field,
$ V(\phi)=\frac{1}{2}\mu^2 \phi^2.$ 
This leads to a more complicated landscape, 
since one more length scale is present.  
Our preliminary numerical results, however,  
suggest that a qualitative similar 
picture to that found in the massless
case.

The situation can be different for other choices of the coupling, $e.g.$ $f(\phi)=\phi^2$, leading to striking  new  features.
For example,  with this coupling,  the phenomenon of spontaneous scalarisation recently discussed in the context of GB scalar-tensor models~\cite{Silva:2017uqg,Doneva:2017bvd,Antoniou:2017acq} (see also~\cite{Herdeiro:2018wub}) should also occur: the Kerr BH is also a solution of the fully
non-linear model (with $\phi=0$), but it may get, in some regions of the parameter space, spontaneously scalarised into a non-GR ECS spinning BH.

\section*{Acknowledgements}
The authors would like to thank V. Cardoso and P. Pani for collaboration in the initial stages of this project.
They also thank L. C. Stein for useful remarks on a first version of the paper. 
C. H. and E.R. acknowledges funding from the FCT-IF programme.  This work was supported by the European  Union's  Horizon  2020  research  and  innovation  programme  under  the H2020-MSCA-RISE-2015 Grant No.   StronGrHEP-690904, the H2020-MSCA-RISE-2017 Grant No. FunFiCO-777740  and  by  the  CIDMA  project UID/MAT/04106/2013. 
The authors  would also  like  to  acknowledge networking support by the COST Action GWverse CA16104. 
Computations were performed
at the Blafis cluster, in Aveiro University.



 \end{document}